\documentclass[amsmath,amssymb,aps,twocolumn]{revtex4-1}

\usepackage{graphicx}
\usepackage{dcolumn}
\usepackage{bm}
\usepackage{hyperref}
\hypersetup{colorlinks=true,linktoc=all,citecolor=blue,linkcolor=blue}
\def\beq{\begin{equation}}
\def\eeq{\end{equation}}
\def\ba{\begin{align}}
\def\ea{\end{align}}
\DeclareMathOperator{\hf}{\frac{1}{2}}

\newcommand{\tcite}[1]{~\cite{#1}}
\newcommand{\braket}[1]{\langle #1 \rangle}

\begin{document}

\title{Can Fermi surface nesting alone drive the charge-density-wave transition in monolayer vanadium diselenide?}

\author{Matthew J. Trott}
\affiliation{SUPA, School of Physics and Astronomy, University of St Andrews, North Haugh, St Andrews, Fife KY16 9SS, United Kingdom}
\author{Chris A. Hooley}
\affiliation{SUPA, School of Physics and Astronomy, University of St Andrews, North Haugh, St Andrews, Fife KY16 9SS, United Kingdom}

\date{13th April 2020}

\begin{abstract}
\noindent
We demonstrate that charge-density-wave formation is possible via a purely electronic mechanism in monolayers of the transition metal dichalcogenide 1T-VSe$_2$. Via a renormalization group treatment of an extended Hubbard model we examine the competition of superconducting and density-wave fluctuations as sections of the Fermi surface are tuned to perfect nesting. We find regions of charge-density-wave order when the Heisenberg exchange interaction is comparable to the Coulomb repulsion, and $d$-wave superconductivity for purely repulsive interactions.  We discuss the possible role of lattice vibrations in enhancing the effective Heisenberg exchange.
\end{abstract}
\maketitle

\noindent
\textit{Introduction.}
Since the isolation and characterization of graphene in 2004\tcite{novoselov2004}, the field of two-dimensional materials has seen an explosion in research activity\tcite{das2015}, and a search has begun for two-dimensional materials that can be tuned to exhibit a wider range of properties than graphene.  Of particular interest in this regard are monolayers of the transition metal chalcogenides FeX (X = Se, Te) and transition metal dichalcogenides MX$_2$ (M = Ti, V, Nb, Mo, Ta, W; X = S, Se, Te)\tcite{manzeli2017}. The transition metal dichalcogenides (TMDs) display an especially wide range of behaviors, including Mott-insulating, semi-metallic, charge-density-wave (CDW), excitonic, and superconducting phases. The development of van der Waals heterostructures made from two or more TMDs\tcite{geim2013} is expected to further increase the range of strongly correlated physics that can be realized in this family of materials.

However, tuning the properties of TMDs requires an understanding of the way in which variations in microscopic parameters affect their phase diagrams.  This, in turn, necessitates an understanding of the physical mechanisms that underlie the experimentally observed ordered phases.  For several of the ordered states of monolayer TMDs, especially the CDW phases, the mechanism remains the subject of debate.

Many TMDs exhibit CDW phases with rather high critical temperatures, which are often further enhanced in the monolayer limit\tcite{yang2014}.  One well known route to CDW formation is via Fermi-surface nesting:\ here sections of the Fermi surface lie parallel to each other, giving an enhanced particle-hole susceptibility at a non-zero wavevector $\textbf{Q}$\tcite{johannes2008,chen2016}. This is an inherently electronic mechanism.  However, there are other candidate mechanisms for the CDW phases in the TMDs, including the softening of phonon modes\tcite{hajiyev2013} and a mechanism based on the transition to an exciton insulator\tcite{rossnagel2011}. 

Here we focus on the 1T structural isomer of vanadium diselenide, VSe$_2$, in the monolayer limit.  Theoretical and experimental attempts to determine the low-temperature Fermi surface of this material do not all agree.  Several studies show column-like Fermi surface pockets protruding from the edge of the Brillouin zone\tcite{zhang2017,esters2017,duvjir2018,chen2018}; others show a Fermi surface with large triangular pockets around the $\text{K}$ and $\text{K}'$ points of the Brillouin zone with an additional small Fermi surface pocket at the $\Gamma$ point\tcite{umemoto2018,feng2018}.  Which of these Fermi surfaces is realized appears to depend on the exact position of the chemical potential with respect to a van Hove singularity in the band structure\tcite{feng2018}.  Such singularities are usually associated with an enhancement of the susceptibilities to various forms of ordered phase, with superconductivity typically dominant\tcite{nandkishore2012,chen2015}.

This variation in the predicted Fermi surface leads to a disagreement over the predicted $\textbf{Q}$-vector of any CDW, and thus also over the reconstructed unit cell. Some studies propose a $\textbf{Q}$-vector perpendicular to the Brillouin zone edge\tcite{mcmillan1975,umemoto2018,sugawara2019}, in the $k_y$ direction; however, others propose alternative nesting vectors parallel to the Brillouin zone edges\tcite{duvjir2018,jang2019}. These studies agree on a renormalization to flat Fermi surface sections in the low-temperature and low-dimensional limit. 

In this article we consider an idealized model of monolayer 1T-VSe$_2$. For definiteness, we assume column-like Fermi surfaces\tcite{duvjir2018,jang2019}, though the patch scheme we employ should also be applicable to the triangular Fermi surface case with appropriate modifications to intra- and inter-pocket scattering and the definitions of superconducting symmetries. We implement a renormalization group (RG) analysis, retaining both particle-particle and particle-hole channels, to capture the interplay of superconducting and density-wave fluctuations, and the effect of Fermi surface nesting on both\tcite{furukawa1998,whitsitt2014}, as the eventual ordered state is approached.

\textit{Model.}
In the low-energy limit we adopt a single-band model to describe the physics of monolayer 1T-VSe$_2$\tcite{duvjir2018,jang2019}. We use an extended Hubbard model, the Hamiltonian of which is given by
\begin{align}\label{ham}
H=&\sum_{\sigma=\uparrow,\downarrow}\sum_{i,j}t_{ij}c^\dagger_{i\sigma}c_{j\sigma}+U\sum_{i}n_{i\uparrow}n_{i\downarrow}\nonumber\\&\qquad\quad\,\,\,\,+\,V\sum_{\sigma\sigma'}\sum_{\braket{i,j}}n_{i\sigma}n_{j\sigma'}+J\sum_{\braket{i,j}}\mathbf{S}_{i}\cdot\mathbf{S}_j.
\end{align}
Here $n_{i\sigma}=c^\dagger_{i\sigma}c_{i\sigma}$ is the number operator for electrons on site $i$ with spin projection $\sigma$, while $\mathbf{S}_i=\hf\sum_{\sigma\sigma'}c^\dagger_{i\sigma}\bm{\tau}_{\sigma\sigma'}c_{i\sigma'}$ is the operator for the spin on site $i$, where $\bm{\tau}=(\tau_x,\tau_y,\tau_z)^\text{T}$ is the vector of Pauli matrices. $t_{ij}$ denotes the hopping matrix elements for our single-band model of VSe$_2$, $U$ and $V$ are the strengths of the on-site and nearest-neighbor parts of the Coulomb repulsion respectively, and $J$ is the Heisenberg exchange coupling. $\braket{i,j}$ indicates that the sum runs over all pairs of nearest-neighbor sites.

\begin{figure}[t]
\begin{center}
\includegraphics[width=0.59\columnwidth]{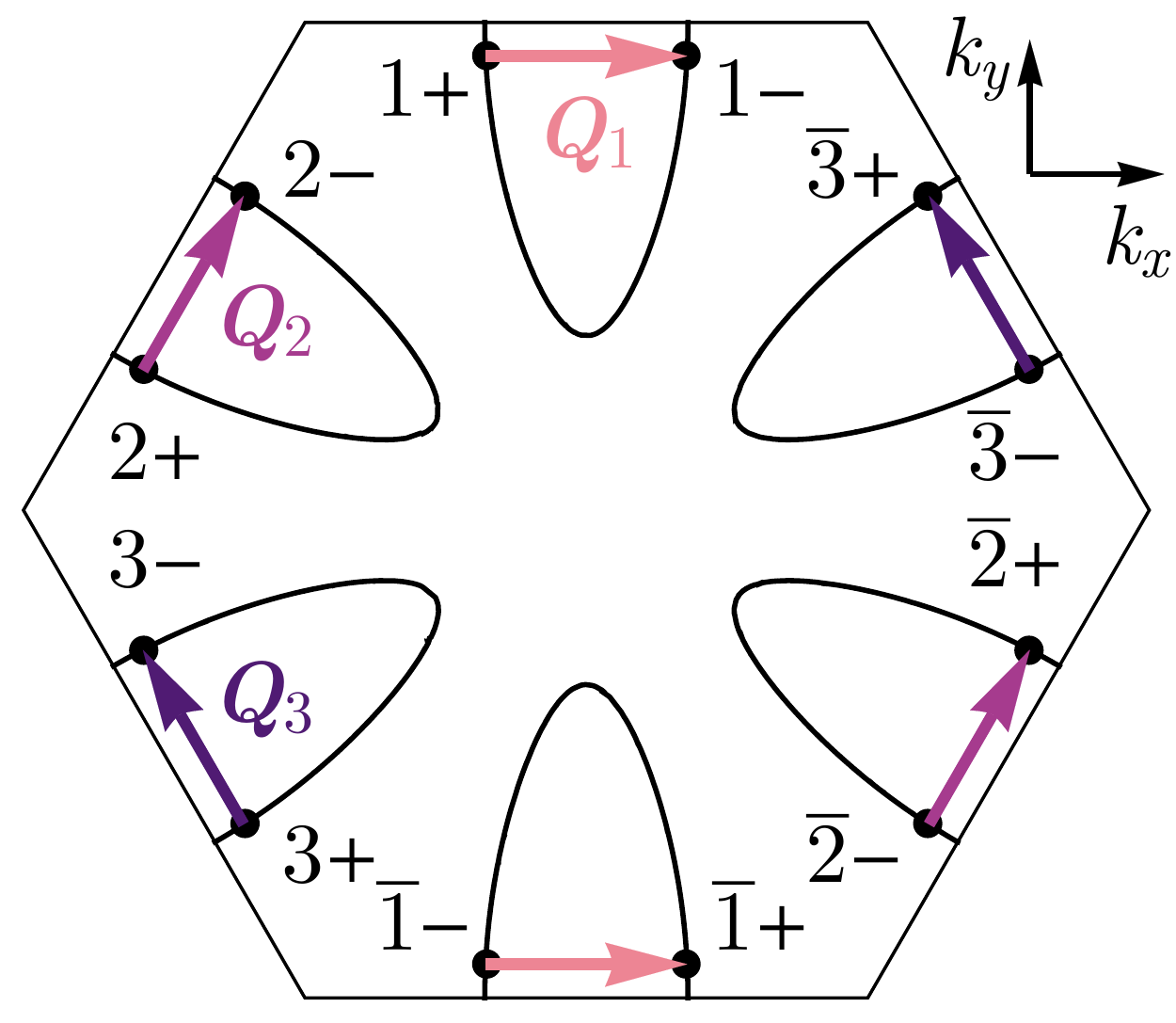}
\includegraphics[width=0.39\columnwidth]{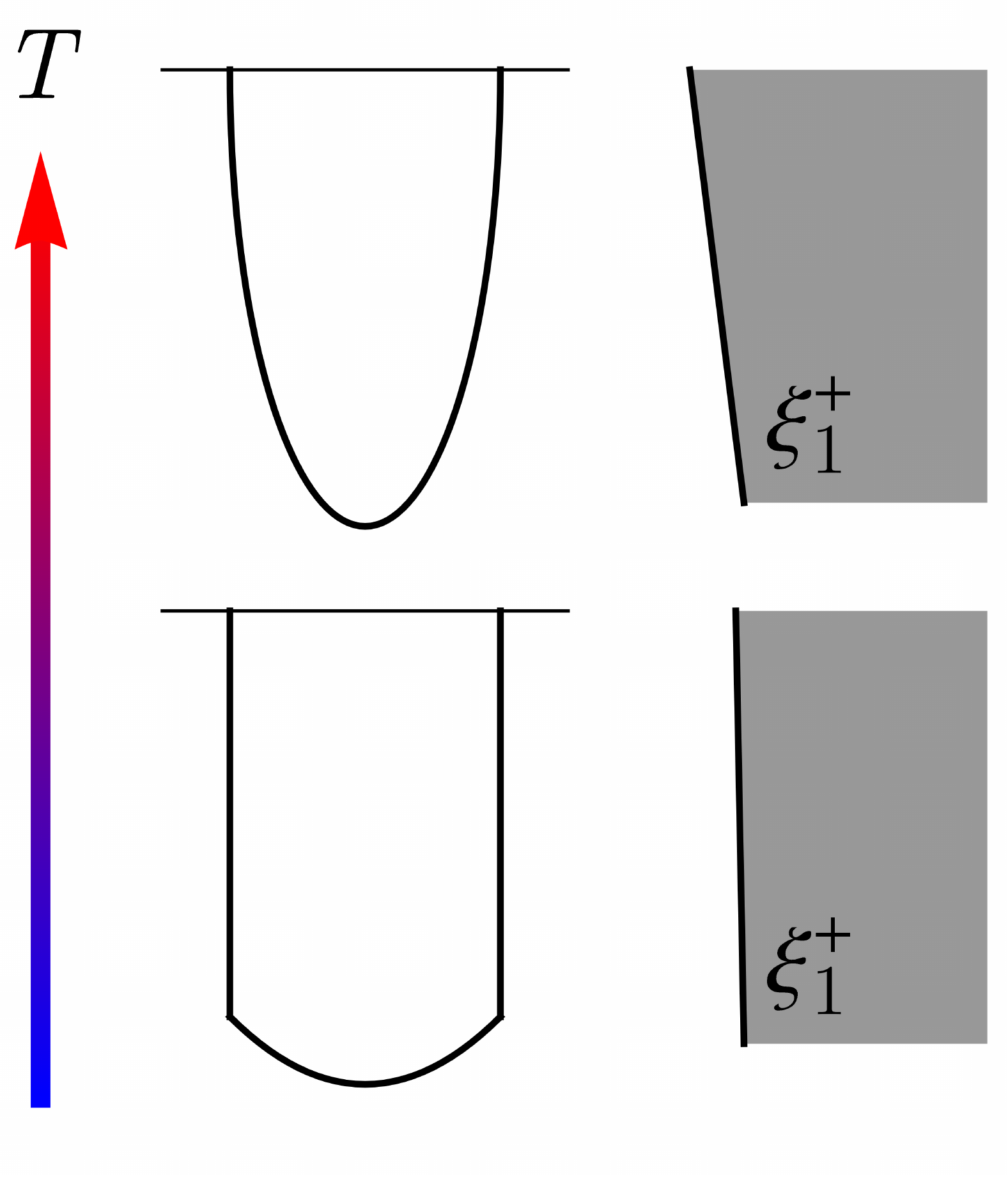}
\end{center}
\caption{Left: Schematic Fermi surface of monolayer 1T-VSe$_2$ with non-nested Fermi surfaces. In the low-temperature limit sections of the Fermi surface become nested. We adopt the notation of Jang \textit{et al.}\tcite{jang2019} to describe the patch scheme and nesting vectors. Right: Schematic change in one pocket of the Fermi surface of monolayer 1T-VSe$_2$ as the temperature $T$ is lowered.  Far right: A zoomed view of the left-hand side of that pocket in our linearized approximation.}
\label{fermis}
\end{figure}

We shall require the form of the non-interacting dispersion relation only near the Fermi energy.  A schematic non-interacting Fermi surface is shown in Fig.~\ref{fermis}. As discussed above, nested sections of the Fermi surface arise at lower temperatures.  Since these dominate the relevant susceptibilities, we can safely use a simplified form of the dispersion relation that agrees with the true dispersion in these nested regions.

We utilize the patch scheme of Jang \textit{et al.}\tcite{jang2019}. This scheme consists of twelve patches that lie on sections of the Fermi surface that become nested at low temperatures, as shown in Fig.~\ref{fermis}. The absolute wavevector of the center of patch $1+$ is denoted $\textbf{K}_{1+}$, and similarly for the other patches.  In each patch we linearize the dispersion relation, i.e.\ we write the single-electron energy (measured with respect to the Fermi energy) as a linear function of the components of $\mathbf{k}$, the wavevector measured relative to the center of the patch.  For the four patches labeled `1', this gives $\xi^{\pm}_{1} =\pm k_x+\varepsilon k_y$ and $\xi^{\pm}_{\overline{1}}=-\xi^{\pm}_1$, in units where both $\hbar$ and the Fermi velocity $v_F$ are set to 1.  The parameter $\varepsilon$ controls the nesting of the Fermi surface, with the limit $\varepsilon \to 0$ corresponding to perfect nesting. We use only the bare dispersions in our calculations as fermion self-energy corrections are independent of the renormalization of interactions at one-loop\tcite{shankar1994}. 

The dispersions on the second and third pockets $n=2,3$ may be obtained from a similar expression, $\xi^{\pm}_{n} \equiv \xi^{\pm}(k^{(n)}_x,k^{(n)}_y)=\pm k^{(n)}_x+\varepsilon k^{(n)}_y$, where the wavevector $\textbf{k}^{(n)}$ is obtained by an appropriate rotation:
\beq 
\begin{pmatrix}
k^{(n)}_x\\k^{(n)}_y 
\end{pmatrix}
=
\begin{pmatrix}
\cos(n-1)\frac{\pi}{3} && \sin(n-1)\frac{\pi}{3}\\ 
-\sin(n-1)\frac{\pi}{3} && \cos(n-1)\frac{\pi}{3}
\end{pmatrix}
\begin{pmatrix}
k_x\\k_y
\end{pmatrix}
,
\eeq
together with the relation $\xi^{\pm}_{\overline{n}}=-\xi^{\pm}_n$.

We can then use these dispersions to calculate the particle-particle and particle-hole susceptibilities for all possible nesting vectors between patches
\begin{eqnarray}
\Pi^{\mathbf{q}}_\text{pp}(\Omega) & = & \int_k G(\omega,\mathbf{k})G(\Omega-\omega,\mathbf{q-k}), \\
\Pi^{\mathbf{q}}_\text{ph}(\Omega) & = & -\int_k G(\omega,\mathbf{k})G(\omega+\Omega,\mathbf{k+q}),
\end{eqnarray}
with $G(\omega,\mathbf{k})=(i\omega-\xi_\mathbf{k}+\mu)^{-1}$. The range of integration is $\omega\in(-\infty,\infty)$ and $k_x,k_y\in(-k_c,k_c)$, where $k_c$ is an ultraviolet momentum cutoff\tcite{whitsitt2014}. 

The complete particle-hole susceptibility at wavevector $\mathbf{Q}_1=\mathbf{K}_{1+}-\mathbf{K}_{1-}$ is
\begin{align}
\Pi^{\mathbf{Q}_1}_\text{ph}(\Omega)=&\frac{k_c}{2\pi ^2 }+\frac{k_c}{4\pi ^2 } \log \left(\frac{\Omega^2+4k_c ^2}{\Omega^2+4\varepsilon^2 k_c ^2}\right)\nonumber\\
&\qquad \qquad \quad -\frac{\Omega}{4 \pi ^2\varepsilon} \arctan\left(\frac{2 k_c  \varepsilon}{\Omega}\right).
\end{align}
The Fermi surface nesting parameter $\varepsilon$ cuts off the $\Omega \to 0$ divergence of the logarithm in this channel, and the height of the $\Omega=0$ peak in the susceptibility reduces as $\varepsilon$ is increased. By contrast, the particle-particle susceptibility at zero momentum in the low-energy limit has the usual logarithmic dependence, independent of $\varepsilon$, $\Pi^\textbf{0}_\text{pp}(\Omega)\approx\frac{k_c}{2\pi^2}\log\left(\frac{k_c}{\Omega}\right)$.  Here we have discarded contributions from non-divergent $\arctan$ terms as they are negligible as $\Pi^0_\text{pp}(\Omega)$ becomes large at low energies.  

The particle-particle susceptibility $\Pi^{\mathbf{q}_1}_\text{pp}(\Omega)$ with $\mathbf{q}_1=\mathbf{K}_{1+}+\mathbf{K}_{1-}$ is logarithmically divergent and dependent on the nesting parameter $\varepsilon$; indeed, $\Pi^{\mathbf{q}_1}_\text{pp}(\Omega)=\Pi^{\mathbf{Q}_1}_\text{ph}(\Omega)$. The particle-hole susceptibility $\Pi^{2\mathbf{K}_{1+}}_\text{ph}(\Omega)$ is always perfectly nested for the case of linear dispersion. However, the nested sections of the VSe$_2$ Fermi surface are finite in length and there will be curvature corrections to the dispersion which will cut off the divergence of the integral. We therefore introduce an additional parameter $\beta$ with $0 \leqslant \beta \leqslant 1$ to reduce the magnitude of this susceptibility and emulate the effect of finite-length nested sections:\ $\Pi^{2\mathbf{K}_{1+}}_\text{ph}(\Omega)=\beta\Pi^0_\text{pp}(\Omega)$.

Interactions between Fermi surface patches belonging to different pockets do not give divergent contributions, since the particle-particle bubble has non-zero $\mathbf{q}$ and there is no particle-hole nesting between patches on separate pockets. Therefore in our low-energy model we retain only one of the Fermi surface pockets, thereby reducing the number of patches to four. This greatly simplifies our effective Lagrangian; however, we lose information about the relative phase of the superconducting order parameter between different Fermi surface pockets and the competition of particle-hole nesting vectors.

\begin{figure}[t]
	\centering
	\begin{tabular}{|c|c|c|}
	\hline
	\includegraphics[width=0.32\columnwidth]{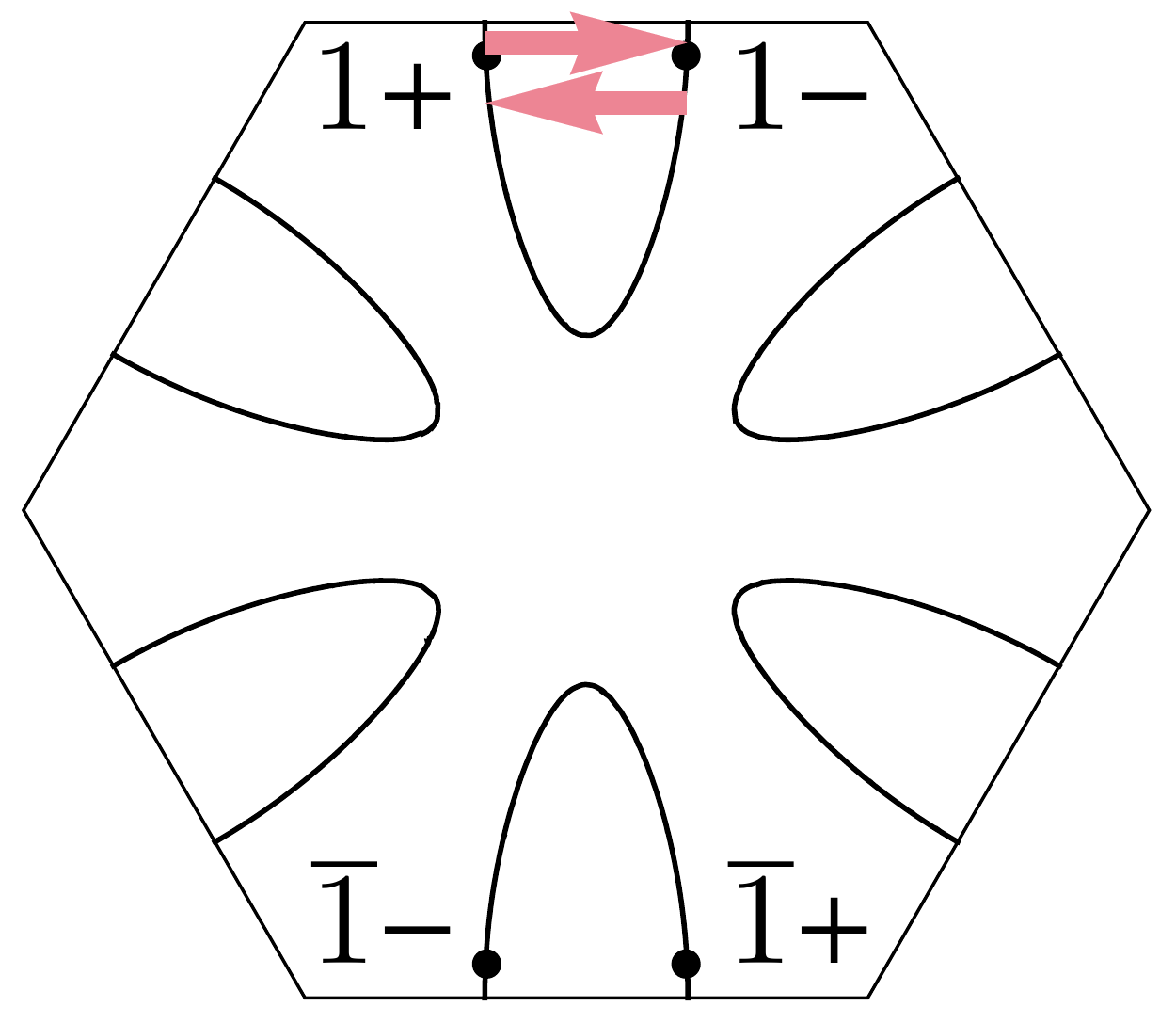}&	
	\includegraphics[width=0.32\columnwidth]{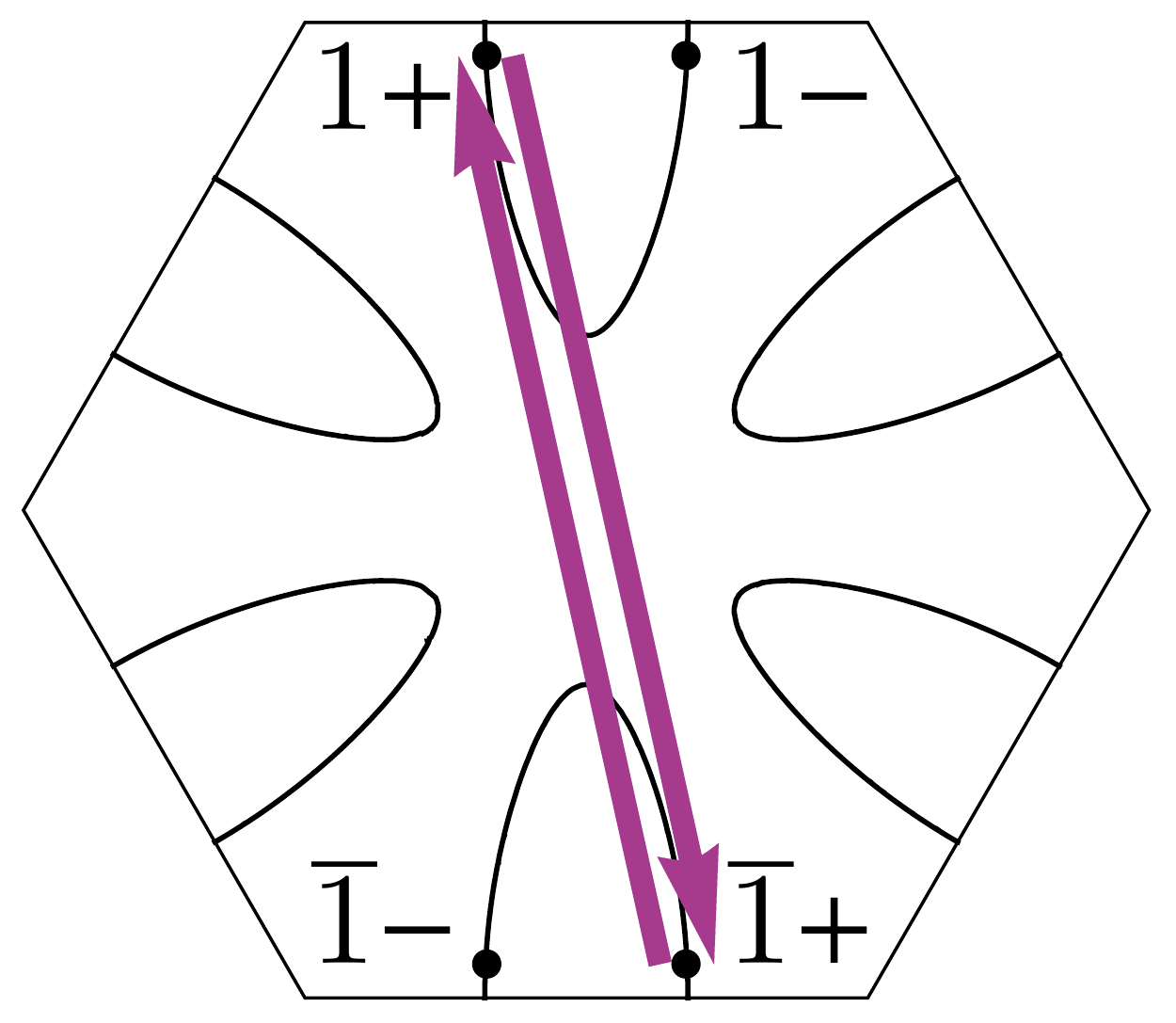}&	
	\includegraphics[width=0.32\columnwidth]{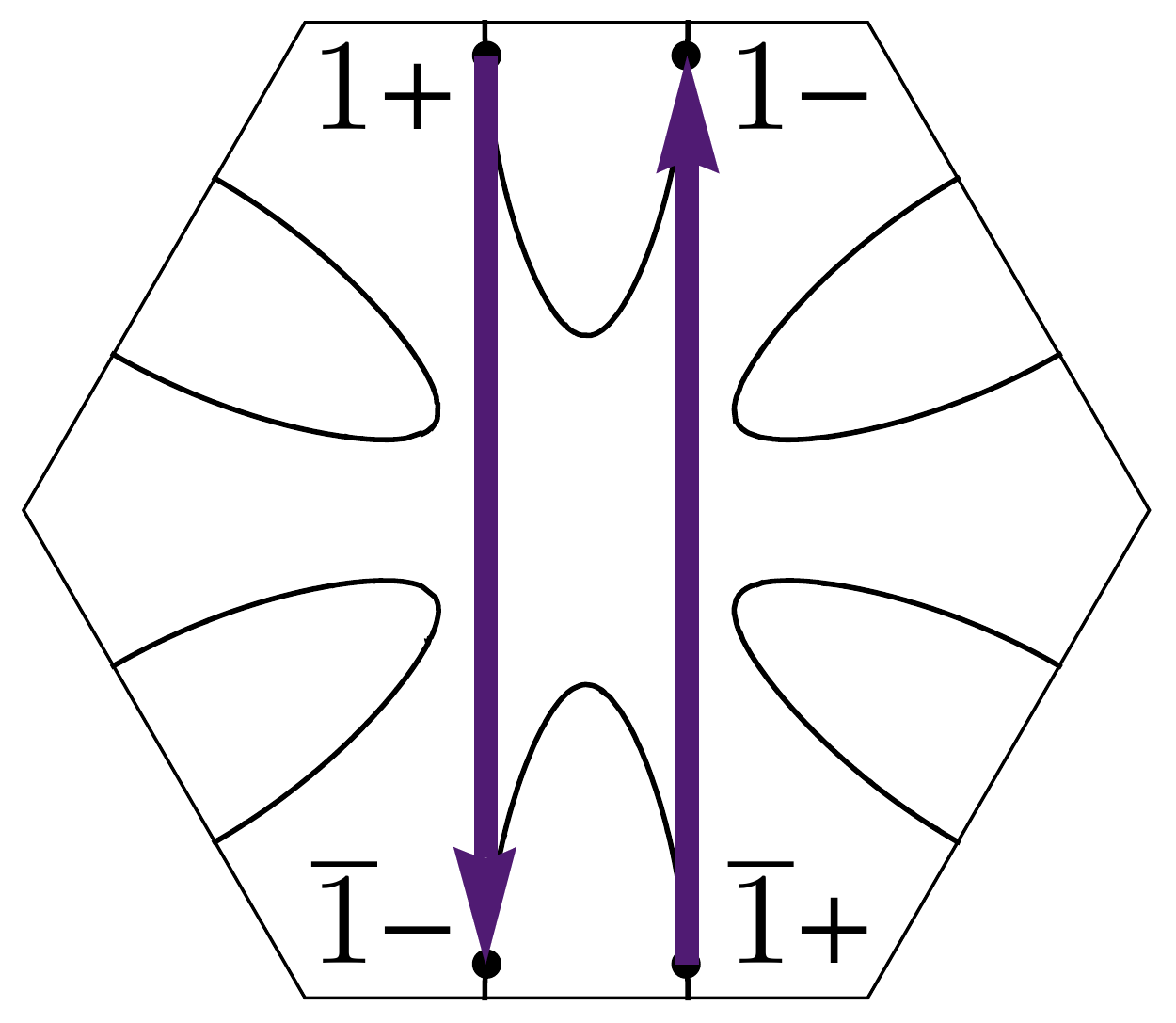}\\
	{$g_1$} & {$g_3$} & {$g_5$}\\
	\hline
	\includegraphics[width=0.32\columnwidth]{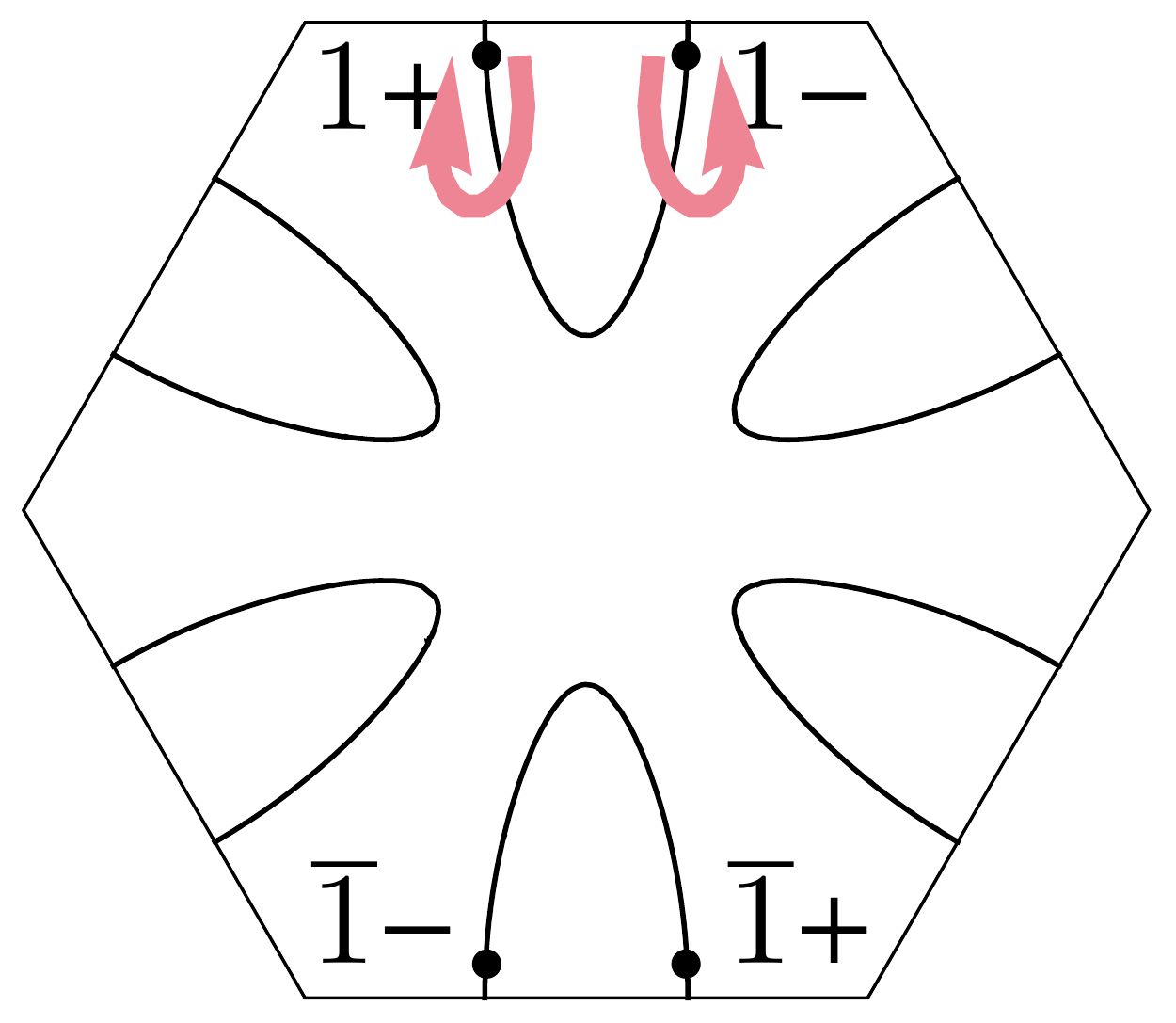}&	
	\includegraphics[width=0.32\columnwidth]{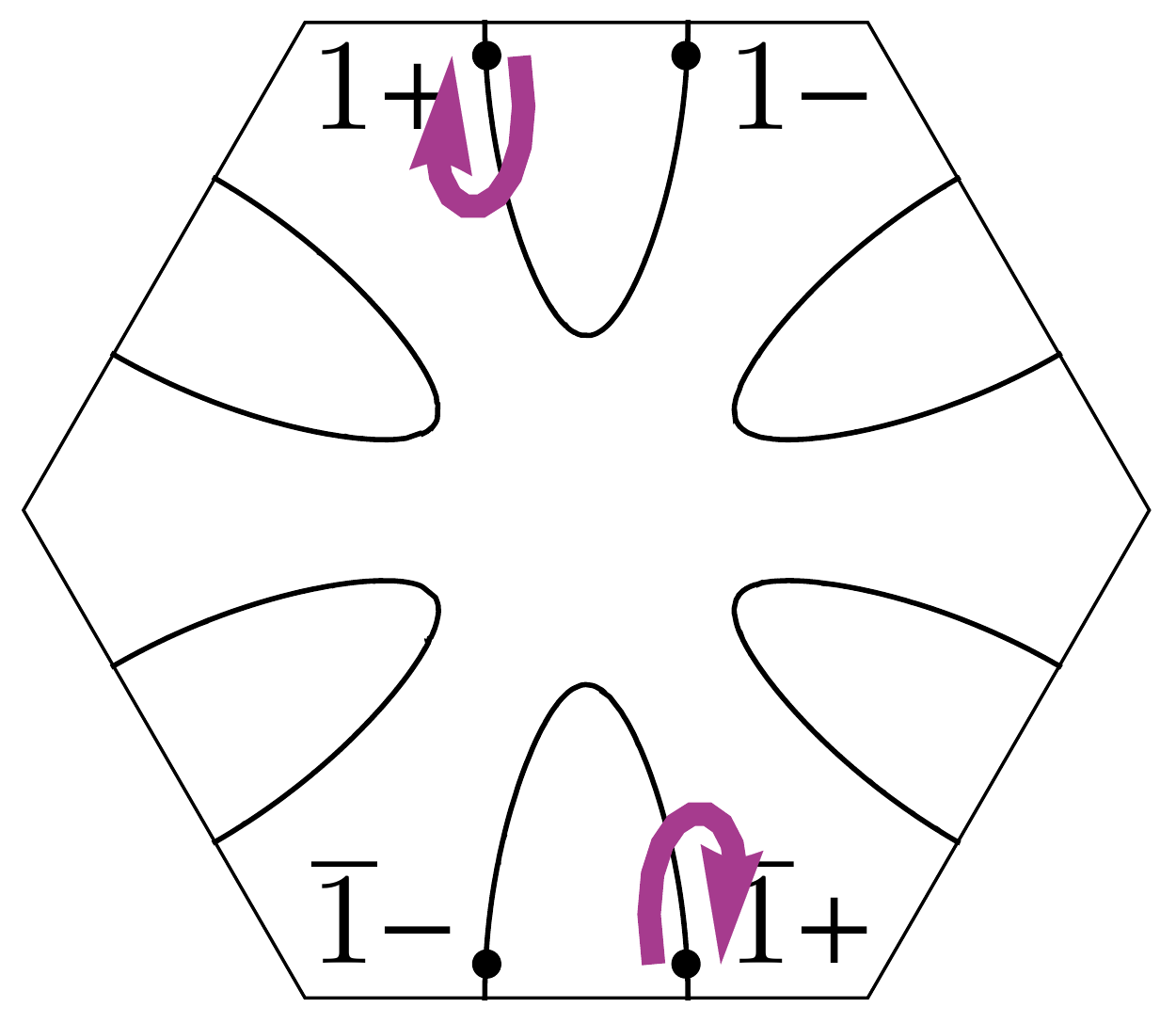}&	
	\includegraphics[width=0.32\columnwidth]{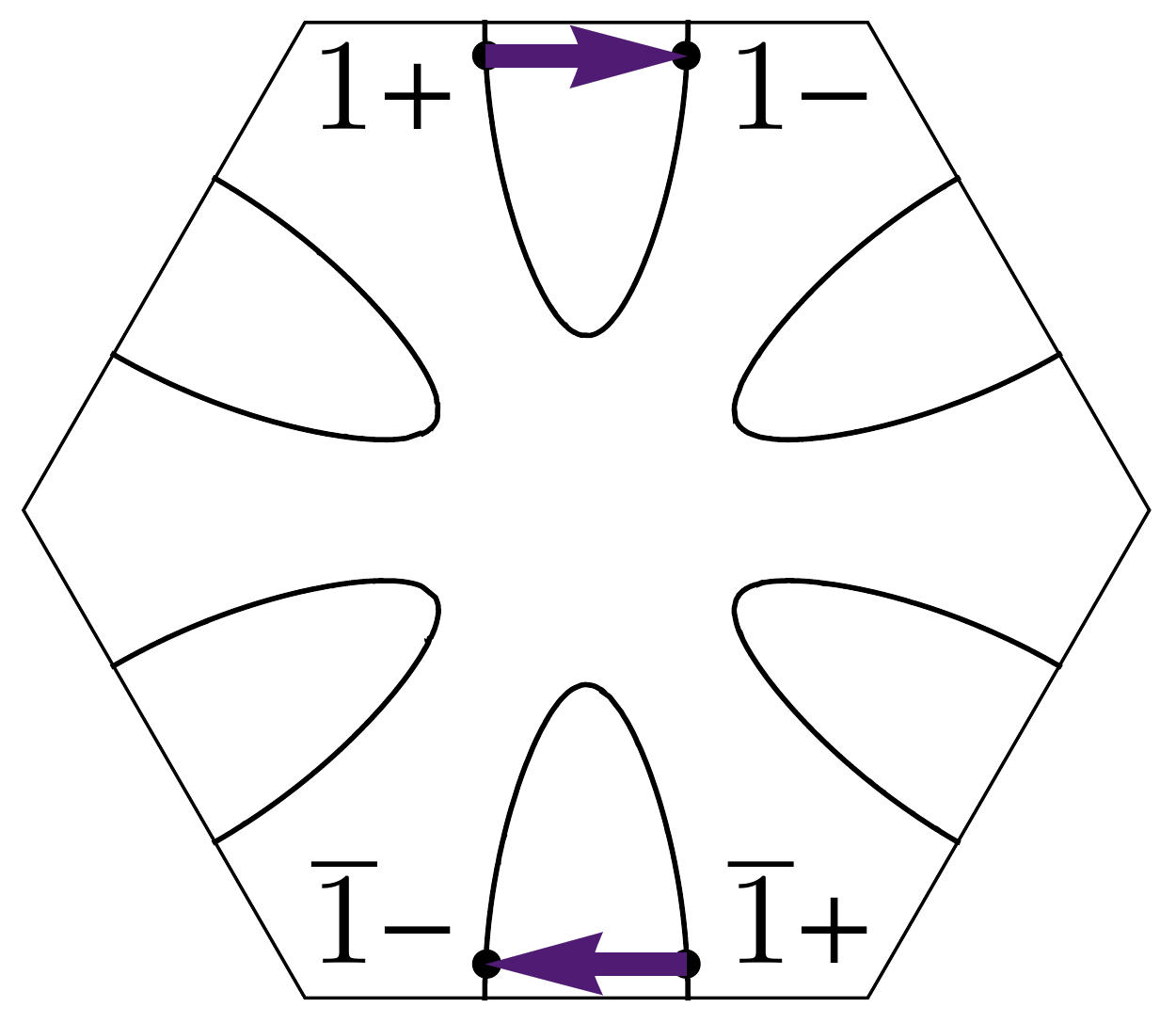}\\
	{$g_2$} & {$g_4$} & {$g_6$}\\
	\hline
	\end{tabular}
	\caption{The six types of two-particle interaction in our low-energy theory of monolayer 1T-VeS$_2$, illustrated for the four patches in group 1.  $g_1$ and $g_2$ are exchange and density-density interactions between patches separated by the wavevector $\mathbf{Q}_1$, $g_3$ and $g_4$ are exchange and density-density interactions between patches with opposite momenta, and $g_5$ and $g_6$ are the two possible exchange interactions that involve all four patches.}
	\label{ints}
\end{figure}

After calculating the divergent susceptibilities, we find that only six of the nine possible interaction terms flow as the theory is renormalized. Retaining only these terms, we obtain the following imaginary-time effective Lagrangian:
\begin{align}
\mathcal{L}=&\sum_{\sigma=\uparrow,\downarrow}\sum_{a=1,\overline{1}}\sum_{s=\pm}\overline{\psi}_{as\sigma}(i\omega-\xi^s_a)\psi_{as\sigma}\nonumber\\
&-g_1\sum_{\sigma\sigma'}\sum_{a}\overline{\psi}_{a+\sigma}\overline{\psi}_{a-\sigma'}{\psi}_{a+\sigma'}{\psi}_{a-\sigma}\nonumber\\
&-g_2\sum_{\sigma\sigma'}\sum_{a}\overline{\psi}_{a+\sigma}\overline{\psi}_{a-\sigma'}{\psi}_{a-\sigma'}{\psi}_{a+\sigma}\nonumber\\
&-\frac{g_3}{2}\sum_{\sigma\sigma'}\sum_{a}\sum_{s}\overline{\psi}_{as\sigma}\overline{\psi}_{\overline{a}s\sigma'}{\psi}_{as\sigma'}{\psi}_{\overline{a}s\sigma}\nonumber\\
&-\frac{g_4}{2}\sum_{\sigma\sigma'}\sum_{a}\sum_{s}\overline{\psi}_{as\sigma}\overline{\psi}_{\overline{a}s\sigma'}{\psi}_{\overline{a}s\sigma'}{\psi}_{as\sigma}\nonumber\\
&-\frac{g_5}{2}\sum_{\sigma\sigma'}\sum_{a}\left[\overline{\psi}_{a+\sigma}\overline{\psi}_{\overline{a}+\sigma'}{\psi}_{a-\sigma'}{\psi}_{\overline{a}-\sigma}+\text{H.c.}\right]\nonumber\\
&-\frac{g_6}{2}\sum_{\sigma\sigma'}\sum_{a}\left[\overline{\psi}_{a+\sigma}\overline{\psi}_{\overline{a}+\sigma'}{\psi}_{\overline{a}-\sigma'}{\psi}_{a-\sigma}+\text{H.c.}\right],
\end{align}
where $\overline{a}$ denote the patch with opposite momentum to $a$. The two-particle scattering processes described by the various interaction terms are shown in Fig.~\ref{ints}.

\textit{Results.} We define the RG flow parameter $y=\log\left(\frac{k_c}{\Omega}\right)$ which diverges to infinity as $\Omega\rightarrow0$. Introducing the dimensionless interactions parameters $g_i\rightarrow\frac{k_c}{2\pi^2}g_i$ we perform a one-loop RG analysis including terms that contribute with a divergent susceptibility at low energies\tcite{furukawa1998,whitsitt2014}. We find the following RG flow equations:
\begin{align}
\dot{g}_1&=2d^{\varepsilon}(y)\left(-g_1^2+g_5g_6-g_6^2\right), \label{diffeq1} \\
\dot{g}_2&=-d^{\varepsilon}(y)g_1^2,\\
\dot{g}_3&=-2\beta g_3^2-2(1-\beta)g_3g_4-2g_5g_6, \\
\dot{g}_4&=-(1-\beta)g_4^2-g_5^2-g_6^2,\\
\dot{g}_5&= -g_3g_6-g_4g_5+2d^{\varepsilon}(y)g_2g_5, \\
\dot{g}_6&= -g_3g_5-g_4g_6+2d^{\varepsilon}(y)(g_1g_5+g_2g_6-2g_1g_6), \label{diffeq6}
\end{align}
where $\dot{g}_i$ denotes the derivative $\frac{dg_i}{dy}$. The $y$-dependence of the couplings $g_i$ has been suppressed for brevity. The function $d^{\varepsilon}(y)$ describes the $\mathbf{Q}_1$ particle-hole susceptibility in terms of the flow parameter $y$ and nesting parameter $\varepsilon$. 

Solving these differential equations, we find that the couplings diverge at a critical value of $y$. In order to allow a numerical solution we stop the flow when the largest of the couplings $g_i$ becomes equal to 1; this defines a critical value $y=y_c$. At this point a subset of the couplings have already become several orders of magnitude larger than their initial values, signalling the breakdown of our perturbation theory and the onset of order. The finite critical value $y_c$ is an artifact of the one-loop RG; higher-loop corrections should shift the divergence to $y_c\rightarrow\infty$. If no coupling has reached 1 by the time $y = 1/U$, we consider no phase transition to occur.

In the limit $y\rightarrow0$ with $\varepsilon$ small,  $d^{\varepsilon}\approx1-\frac{\varepsilon^2}{3\log2}$. In the large-$y$ limit $d^{\varepsilon}(y)$ takes the form $d^{\varepsilon}(y\rightarrow y_c)=((1-\varepsilon^2)e^y)/((1+e^y)(1+\varepsilon^2e^y))$. We therefore use the following approximation to $d^{\varepsilon}(y)$:
\beq\label{dfun}
d^{\varepsilon}(y)=\frac{d\Pi^{\mathbf{Q}_1}_\text{ph}}{d\Pi^0_\text{pp}}\approx\frac{1-\frac{\varepsilon^2}{3\log2}}{1+\varepsilon^2e^y},
\eeq
which interpolates between the $y\rightarrow0$ and $y\rightarrow y_c$ limits.

\begin{figure*}[t]
	\centering
	\includegraphics[width=0.98\textwidth]{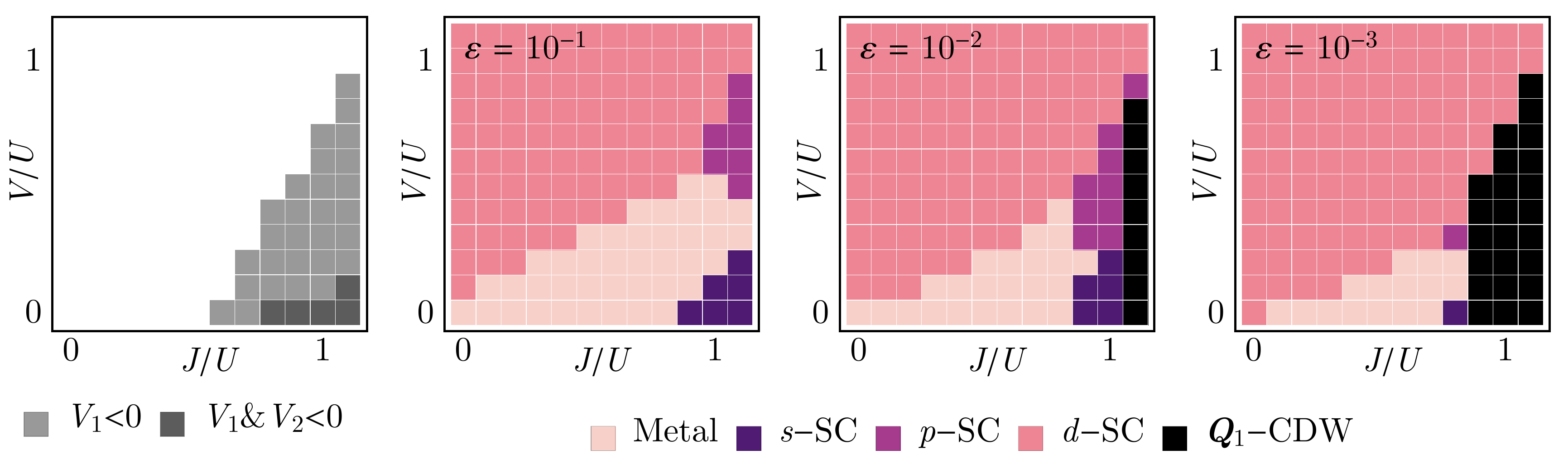}
	\caption{Leftmost panel: An illustration of the mapping between the nearest-neighbor Coulomb repulsion $V$ and Heisenberg exchange interaction $J$ and the coupling constants $V_1$ and $V_2$, defined in (\ref{v12}). Other panels: Calculated phase diagrams for our model of monolayer 1T-VSe$_2$.  The assumed degree of Fermi surface nesting increases from left to right ($\varepsilon=10^{-1},10^{-2},10^{-3}$), and the parameter $\beta$ that represents the finite length of the nested sections is set to $1/2$.}
	\label{phases}
\end{figure*}

The initial conditions for the couplings are approximated by
\begin{align}
V_1\approx U+V-\frac{7}{4}J, &&
V_2\approx U+3V-\frac{5}{4}J.  \label{v12}
\end{align}
We find the initial conditions at $y=0$ for the couplings to be $g_1^0\approx g_3^0\approx g_6^0\approx V_1$ and $g_2^0\approx g_4^0\approx g_5^0\approx V_2$ in our approximation. The effect of this approximation is to split the solutions into three regions:\ (i) $V_1, V_2 > 0$, all couplings repulsive in the ultraviolet; (ii) $V_2 > 0, V_1 < 0$; and (iii) $V_1, V_2 < 0$, i.e.\ all couplings attractive.  In a more general microscopic model the values of the couplings $g_i^0$ would be independent.  The mapping between the microscopic couplings $J$ and $V$ and the RG couplings $V_1$ and $V_2$ is illustrated in Fig.~\ref{phases}.

Additionally we must calculate the susceptibilities of the possible order parameters. We therefore introduce test vertices for all possible two-particle correlators and calculate the corresponding one-loop vertex corrections. In the particle-particle channel the eigenvectors $\Delta_s=\Delta(1,1,1,1)^\text{T}/2$, $\Delta_d=\Delta(-1,1,-1,1)^\text{T}$/2, $\Delta_p=\Delta(-1,-1,1,1)^\text{T}$/2, and $\Delta_f=\Delta(1,-1,-1,1)^\text{T}/2$ define the pairing symmetry.  The corresponding eigenvalues are given in equations (\ref{ssc}--\ref{fsc}).  The SDW and CDW susceptibilities are calculated via $\chi^\mathbf{q}_\text{SDW}=\chi_{\uparrow\uparrow}^{\mathbf{q},\text{ph}}-\chi_{\downarrow\downarrow}^{\mathbf{q},\text{ph}}$, $\chi^\mathbf{q}_\text{CDW}=\chi_{\uparrow\uparrow}^{\mathbf{q},\text{ph}}+\chi_{\downarrow\downarrow}^{\mathbf{q},\text{ph}}$\tcite{whitsitt2014}. 

We refer to the possible superconducting symmetries using their continuum analogs, despite the fact that our system is on a lattice and we furthermore only utilize a discrete set of patches.  To make the meanings of these order parameters clear, we note that the $s$-wave eigenvector predicts an isotropic gap, while the $d$-wave eigenvector leads to four nodes on each Fermi surface pocket.  The $p$-wave and $f$-wave eigenvectors each give two nodes per pocket; however, the $p$-wave order parameter na{\"\i}vely changes sign twice as a function of angle in the Brillouin zone, whereas the $f$-wave order parameter changes sign six times.

Due to our patch approximation we can predict neither the relative phases of the superconducting order parameter between pockets nor which vector(s) $\mathbf{Q}_i$ will form the CDW.  To calculate the latter, a multi-component order parameter theory is required\tcite{jang2019}.

Given the divergence of the couplings at $y_c$ we introduce the asymptotic form $g_i=G_i/(y_c-y).$ As $y\rightarrow y_c$ we can express the divergences of order parameter susceptibilities in the power-law form $\chi_j=(y_c-y)^{-j}$, with $j\in$\,\,\{$\alpha^s_{\text{SC}}$, $\alpha^d_{\text{SC}}$, $\alpha^p_{\text{SC}}$, $\alpha^f_{\text{SC}}$, $\alpha^{\mathbf{Q}_1}_{\text{SDW}}$, $\alpha^{\mathbf{Q}_1}_{\text{CDW}}$, $\alpha^{2\mathbf{K}_{1+}}_{\text{SDW}}$, $\alpha^{2\mathbf{K}_{1+}}_{\text{CDW}}$\}. The exponents are given by the following equations:
\begin{align}
\alpha^s_{\text{SC}}&= -G_3-G_4-G_5-G_6,\label{ssc}\\
\alpha^d_{\text{SC}}&=-G_3-G_4+G_5+G_6,\\
\alpha^p_{\text{SC}}&=G_3-G_4+G_5-G_6,\\
\alpha^f_{\text{SC}}&=G_3-G_4-G_5+G_6\label{fsc},\\
\alpha^{\mathbf{Q}_1}_{\text{SDW}}&= d^{\varepsilon}(y_c)\left(G_2+G_5\right), \\
\alpha^{\mathbf{Q}_1}_{\text{CDW}}&= d^{\varepsilon}(y_c)\left(-2G_1+G_2+G_5-2G_6\right), \\
\alpha^{2\mathbf{K}_{1+}}_{\text{SDW}}&= \beta G_4,\\
\alpha^{2\mathbf{K}_{1+}}_{\text{CDW}}&= \beta\left(G_4-2G_3\right).
\end{align}
Due to the nature of our patch scheme, ferromagnetic instabilities cannot be investigated:\ they require the full Fermi surface to calculate the susceptibilities. Ferromagnetic phases have been observed experimentally in monolayer VSe$_2$\tcite{bonilla2018}. However, there is evidence to suggest that ferromagnetism is suppressed near the CDW phase as our nested approximation would suggest\tcite{fumega2019}.

Considering the alternative triangular Fermi surface case, the definitions of intra- vs. inter-pocket scattering have to be altered. This does not change the CDW nesting vectors; however, the $f$-wave superconductivity would be replaced by an $s_{\pm}$-like order parameter.

Solving (\ref{diffeq1}--\ref{diffeq6}) numerically with the initial conditions $g_i(y=0)=g^0_i$, and utilizing the definitions of the divergent susceptibilities, we can investigate the phase diagram of the model. In the case of a pure contact interaction, for which $V=J=0$, only two instabilities are predicted:\ $s$-wave superconductivity for an initially attractive interaction and $d$-wave superconductivity for an initially repulsive one. For $V$ and $J$ non-zero, the phase diagrams for a range of nesting strengths ($\varepsilon=10^{-1},10^{-2},10^{-3}$) are plotted in Fig.~\ref{phases}. When all interactions are initially repulsive the predicted instability is again to $d$-wave superconductivity. As some of the initial interactions become attractive, regions of $s$-wave and $p$-wave superconductivity arise. As the nesting strength is increased, these regions become occupied by a CDW phase.

\textit{Summary and discussion.}
To analyze the effect of Fermi surface nesting on CDW formation in the TMDs, we have performed an RG analysis of an extended Hubbard model for monolayer 1T-VSe$_2$, retaining both particle-particle (superconducting) and particle-hole (density wave) channels. In the region of parameters where some, or all, of the bare two-particle interactions are attractive, regions of superconductivity give way to CDW order as the strength of Fermi surface nesting is increased.

The tuning of the Fermi surface nesting is a control parameter in our analysis. However, taking into account self-energy corrections to the patch dispersions, a flow to perfect nesting is predicted by previous RG calculations\tcite{jang2019,metlitski2010,sur2015,sur2016}. Our analysis is therefore complementary to that of Jang \textit{et al.}\tcite{jang2019}, and predicts CDW formation without any mean-field assumption, taking into account the competition of superconducting and density wave fluctuations.

The fact that some of the bare interactions should be attractive for a CDW phase to be favored is an interesting result in the context of a purely electronic calculation. It is well known that electron-phonon interactions lead to an effective attractive interaction between electrons. The result would suggest that additional phonon effects could replace or coexist with the role of exchange interaction and further enhance the CDW phase. 

The $\mathbf{Q}_1$ CDW wavevector is favored as the chosen instability, even with the artificial enhancement of the $\mathbf{q}_1$ channel due to lack of curvature corrections to the dispersions. Thus this behavior again agrees with that of Jang \textit{et al.}\tcite{jang2019} and gives a viable prediction for a nesting mechanism in the monolayer TMDs.

\textit{Acknowledgments.} MJT acknowledges financial support from the CM-CDT under EPSRC (UK) grant number EP/L015110/1. CAH acknowledges financial support from the EPSRC (UK), grant number EP/R031924/1.

\end{document}